\documentclass[useAMS,usenatbib,]{mn2e}
\usepackage{rotating}

\newcommand{\be} {\begin{equation}}

\newcommand{\ee} {\end{equation}}

\newcommand{\RXTE}{{\em R}XTE}
\newcommand{\CXO}{{\em Chandra}}
\newcommand{\bc}{\begin{center}}
\newcommand{\ec}{\end{center}}

\def\ltsima{$\; \buildrel < \over \sim \;$}
\def\lsim{\lower.5ex\hbox{\ltsima}}
\def\loe{\lower.5ex\hbox{\ltsima}}
\def\gtsima{$\; \buildrel > \over \sim \;$}
\def\gsim{\lower.5ex\hbox{\gtsima}}
\def\goe{\lower.5ex\hbox{\gtsima}}

\def \cm2{cm$^{-2}$\,}

\def\ergs {erg\,s$^{-1}$}
\def\ergscm2 {erg\,s$^{-1}$cm$^{-2}$}

\def\srca{LS\,I\,+61$^{\circ}$303}
\def\srcb{LS\,5039\,}

\begin{document}

\title[Deep Chandra observations of \srcb]{Deep \CXO\, observations of TeV binaries - II: \srcb}

\author[Rea et al.]{N. Rea$^{1}$\thanks{Ramon y Cajal Research Fellow; rea@ieec.uab.es.}, D.~F. Torres$^{1,2}$, G. A. Caliandro$^{1}$,  D. Hadasch$^{1}$, M. van der Klis$^{3}$,  \newauthor P.~G. Jonker$^{4,5,6}$, M. M\'endez$^{7}$, A. Sierpowska-Bartosik$^{8}$\\
$^{1}$ Institut de Ci\'encies de l'Espai (IEEC--CSIC), Campus UAB, Fac. de Ci\'encies, Torre C5-parell, 2a planta, 08193 Barcelona, Spain \\
$^{2}$ Instituci\'o Catalana de Recerca i Estudis Avan\c{c}ats (ICREA),
Barcelona, Spain \\
$^{3}$ University of Amsterdam, Astronomical Institute ``Anton Pannekoek'', Postbus 94249, 1090 GE, Amsterdam, The Netherlands \\
$^{4}$ SRON-Netherlands Institute for Space Research, Sorbonnelaan 2, 3584 CA, Utrecht, the Netherlands \\
$^{5}$ Harvard-Smithsonian Center for Astrophysics, 60 Garden Street, Cambridge, MA 02138, USA \\
$^{6}$ Department of Astrophysics, IMAPP, Radboud University Nijmegen, Heyendaalseweg 135, 6525 AJ, Nijmegen, The Netherlands\\
$^{7}$ Kapteyn Astronomical Institute, University of Groningen, PO Box 800, 9700 AV, Groningen, The Netherlands\\
$^{8}$ University of Lodz, Department of Astrophysics, Pormorska street 149/153, PL-90236, Lodz, Poland 
}

\input psfig.sty

\pagerange{\pageref{firstpage}--\pageref{lastpage}} \pubyear{2009}

\maketitle

\label{firstpage}

\begin{abstract}

We report on  \CXO\, observations of the TeV emitting High Mass
X--ray Binary \srcb, for a total exposure of $\sim$70\,ks, using the ACIS-S camera in Continuos Clocking
mode to search for a possible X-ray pulsar in this system.  We did not
find any periodic or quasi-periodic signal in the  0.3--0.4 and 0.75--0.9 orbital phases, and in
a frequency range of $0.005-175$\, Hz. We derived an average pulsed
fraction 3$\sigma$ upper limit for the presence of a periodic signal
of $\lsim$15\% (depending on the frequency and the energy band), the
deepest limit ever reached for this object.  If the X-ray emission of \srcb\, is due (at least in part) to a rotational powered pulsar, the latter is either spinning faster than $\sim5.6$\,ms, or having a beam pointing away from our line of sight, or contributing to $\lsim$15\% of the total X-ray emission of the system in the orbital phases we observed.

\end{abstract}

\begin{keywords}
X-ray: binaries -- (stars:) binaries: individual: \srcb
\end{keywords}

\section{Introduction}

The characterization of the compact objects hosted in the two
TeV	binaries \srca\, and \srcb\,  is now one long-standing question in high-energy astrophysics.
The peculiar GeV (Abdo et al. 2009a) and TeV (Aharonian et al. 2005, 2006) emission of these two High Mass X-ray Binaries (HMXBs) has been modeled as being powered either by the interaction of pulsar-accelerated particles in the pulsar wind zone (e.g., Sierpowska-Bartosik \& Torres 2008) and/or by the interaction of shock-accelerated particles driven by the collision between the winds of the pulsar and the massive companion star (e.g., Dubus 2006); or by a jet emitted by an accreting black-hole or neutron star (e.g., Khangulyan et al. 2008).  No firm conclusion on this issue has been drawn thus far and the literature contains detailed models and discussions on all of these alternatives.


\begin{figure*}
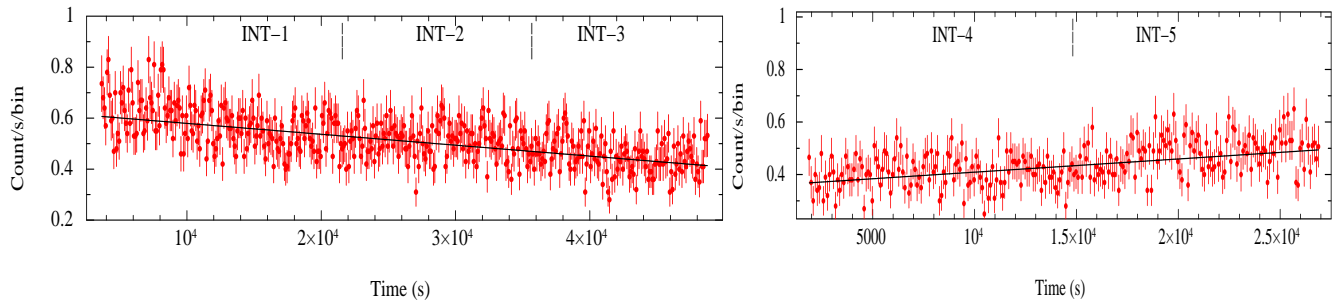

\hbox{\psfig{figure=lightcurve_100s_1_colin_int.ps,width=9.5cm,height=4cm,angle=270}
\psfig{figure=lightcurve_100s_2_colin_int.ps,width=8.0cm,height=4cm,angle=270}}
\caption{\CXO\, background-subtracted 0.3--10\,keV lightcurves of the two segments of the \CXO\, observation of \srcb\, binned at 100\,s. }
\label{lcurve}
\end{figure*}


In Rea et al. (2010; Paper I) we have reported on a $\sim$ 95 ks Chandra observation of  \srca\, aimed at gathering the most stringent limit on (or detecting) the pulsed fraction of any X-ray signal from a putative energetic pulsar. In that paper we inferred an X-ray pulsed fraction for  \srca\, of  less than $10\%$ (depending on the frequency and on the energy band), while the compact object was close to the apastron of its $\sim$26 days orbit around the Be companion. We also observed two flares from \srca\, which showed a harder spectrum with respect to the source emission, and can be explained in an accretion scenario by variability in the accretion rate, while in the rotational-powered pulsar one, as the interaction between the pulsar wind and clumps in the Be wind. Similar flares have been observed by other experiments, among them, by RXTE (Smith et al. 2009, Torres et al. 2010, Li et al. 2011) and Swift-XRT (Esposito et al. 2006). \\
The variation of the spectral index with flux has also been found for \srcb\, (Bosch-Ramon et al. 2005, Takahashi et al. 2009). However, the steadiness of the lightcurve is unique to this latter source, whereas there is significant orbit-to-orbit variability in  \srca\,  (Torres et al. 2010, Li et al. 2011). Kishishita et al. (2009) reported that \srcb\, has a X-ray lightcurve presenting long-term stability using multi-satellite observations between 1999 and 2007\footnote{We note that Durant et al. (2011), using new {\em Chandra} observations, report a flux $\sim$2 times lower , and a smaller photon-index, than that observed by {\em Suzaku} (Kishishita et al. 2009) at the same orbital phase. This finding might represent a hint for a long-term variability, and they contradict the well-settled flux-gamma anti-correlation with the orbital phase (i.e. low flux always corresponded to large photon-indexes). We caveat  that assuming the flux and spectrum of the {\em Suzaku} observation for the orbital phase covered by the {\em Chandra} one, we find that the pile-up in {\em Chandra} should be $>20$\%, and this might in principle account for the lower flux and harder photon index.}.

With these prior results we would not  a priori expect to find in our Chandra observations anything different from this regarding the lightcurve evolution of LS 5039; our aims being rather the search for accretion and pulsation signatures.
One possible explanation for the distinction in the lightcurve evolution of LS 5039 and \srca\,  is the fact that  whereas the former system
host an O star, \srca\,  has  a Be star primary, which is characterized with a dense equatorial disk. If, for instance, there are disc structure changes in time, the latter may induce variability in X-ray modulation, deviating its behavior for the clock-like emission found for \srcb  .

The detection of fast X-ray pulsations would be an unambiguous tracer for a secure determination of the neutron star nature of the compact objects hosted in TeV binaries, if such is the case. Of course, in the accretion scenario 
Type I X-ray bursts might be another neutron star tracer, although never observed 
so far in a HMXB given the relatively high magnetic field of these young objects. In fact,  for magnetic fields larger than $\sim10^{10}$\,G the thermo-nuclear instability giving rise to Type I X-ray bursts is suppressed, as the accreted H and He is burned stably in the accretion column of high B-field neutron stars (Lewin et al. 1995).
Searches for pulsations from a young neutron star in massive binaries have in 
general many more chances of success in the X-ray than in
the radio band. In fact, if the putative young neutron star is accreting, radio pulsed 
emission is expected to be halted by the accretion process, while if not accreting a) 
the X-ray pulsar beam is larger than the radio one, and b) the strong companion wind 
might well prevent the detection of radio pulsations because of free-free absorption and the 
large and highly variable Dispersion Measure (DM) induced by the
wind (see e.g., Zdziarski et al. 2008).


\begin{figure*}
\hbox{
\psfig{figure=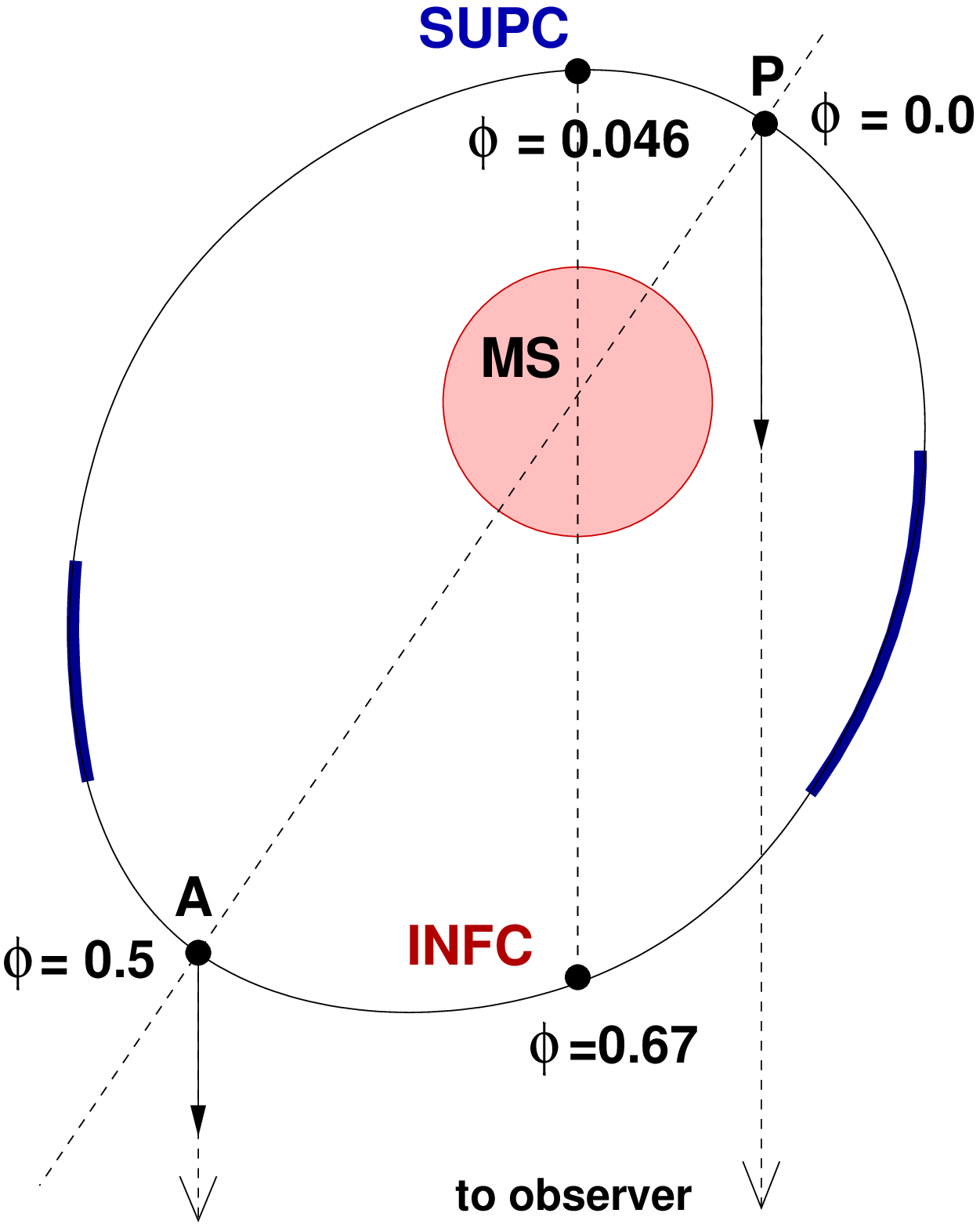,width=6cm}
\psfig{figure=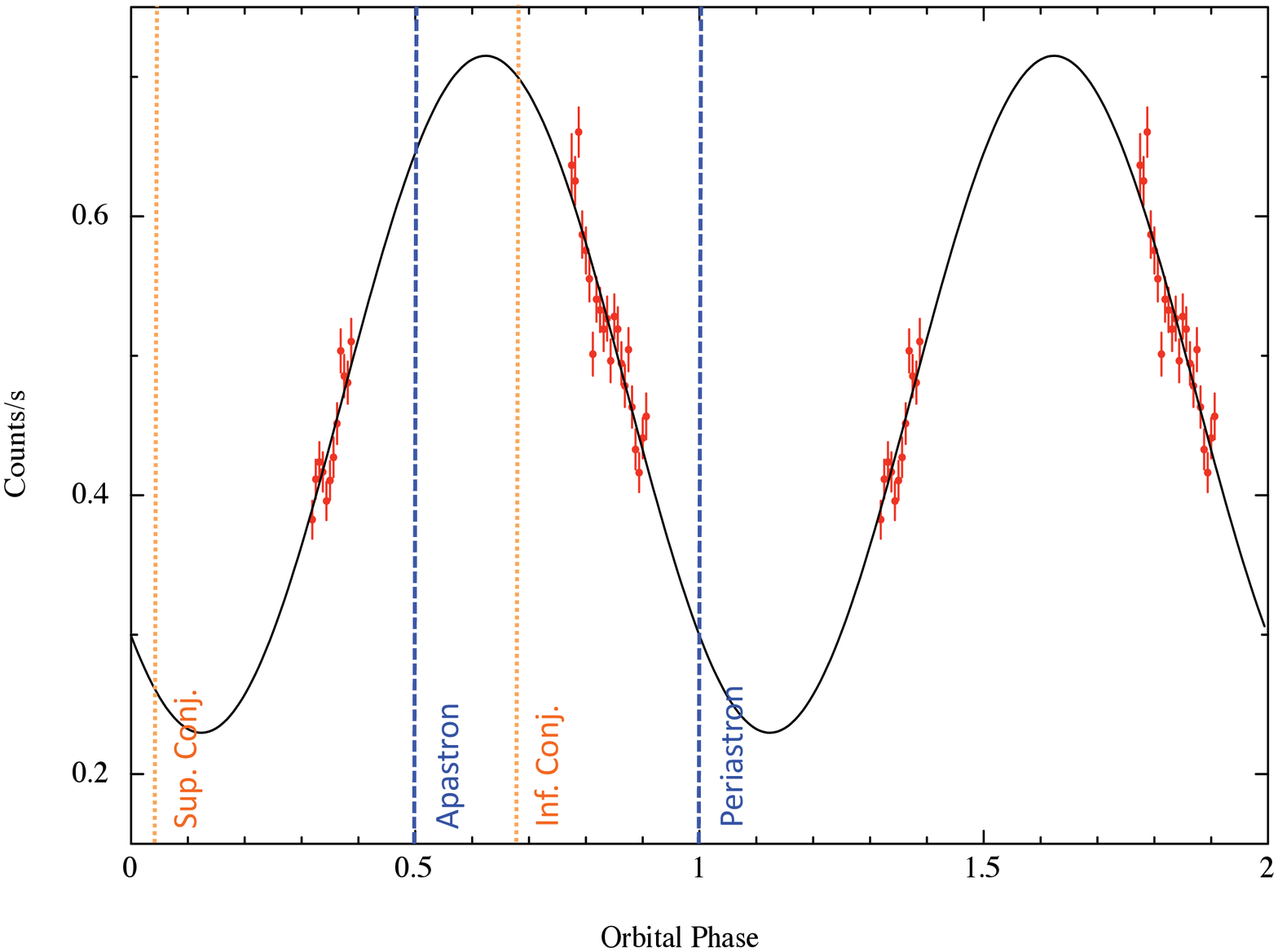,width=11cm,height=7cm}}
\caption{{\em Left panel}: \srcb's geometry considering the orbital solution of Aragona et al. (2009), and the phases for Inferior conjunction (INFC), Superior conjunction (SUPC), periastron (P), and apastron (A) are marked
accordingly, but the inclination is not taken into account in the plot. The orbit (black solid line) and the massive star (MS; in orange) are roughly to scale. The blue thick line indicates the orbital phases spanned by our two \CXO\, observations. {\em Right panel}: \CXO\, data folded with the orbital ephemeris of \srcb .}
\label{orbit}
\end{figure*}



\begin{figure}
\vbox{
\psfig{figure=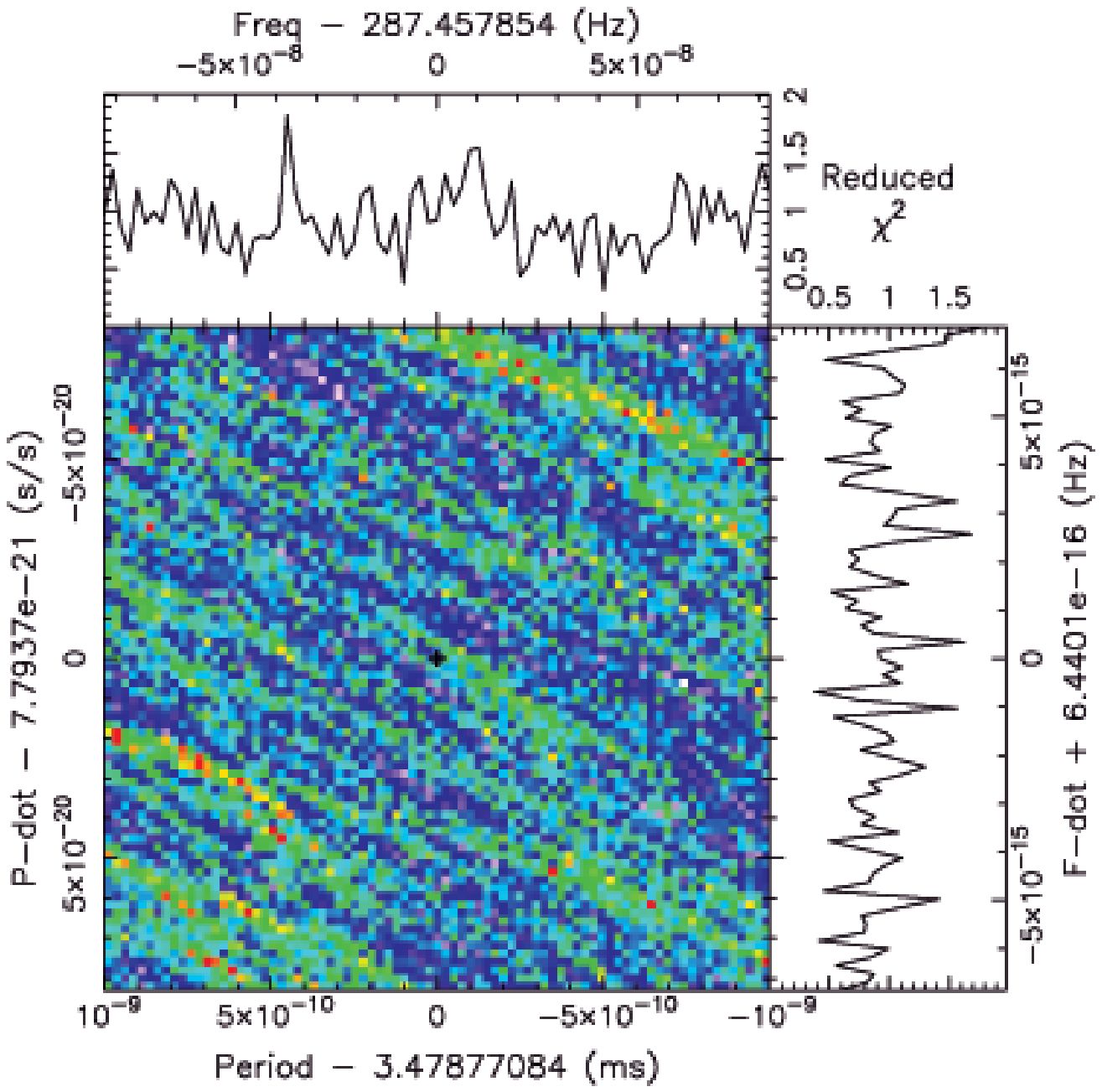,width=8cm,height=6cm}
\psfig{figure=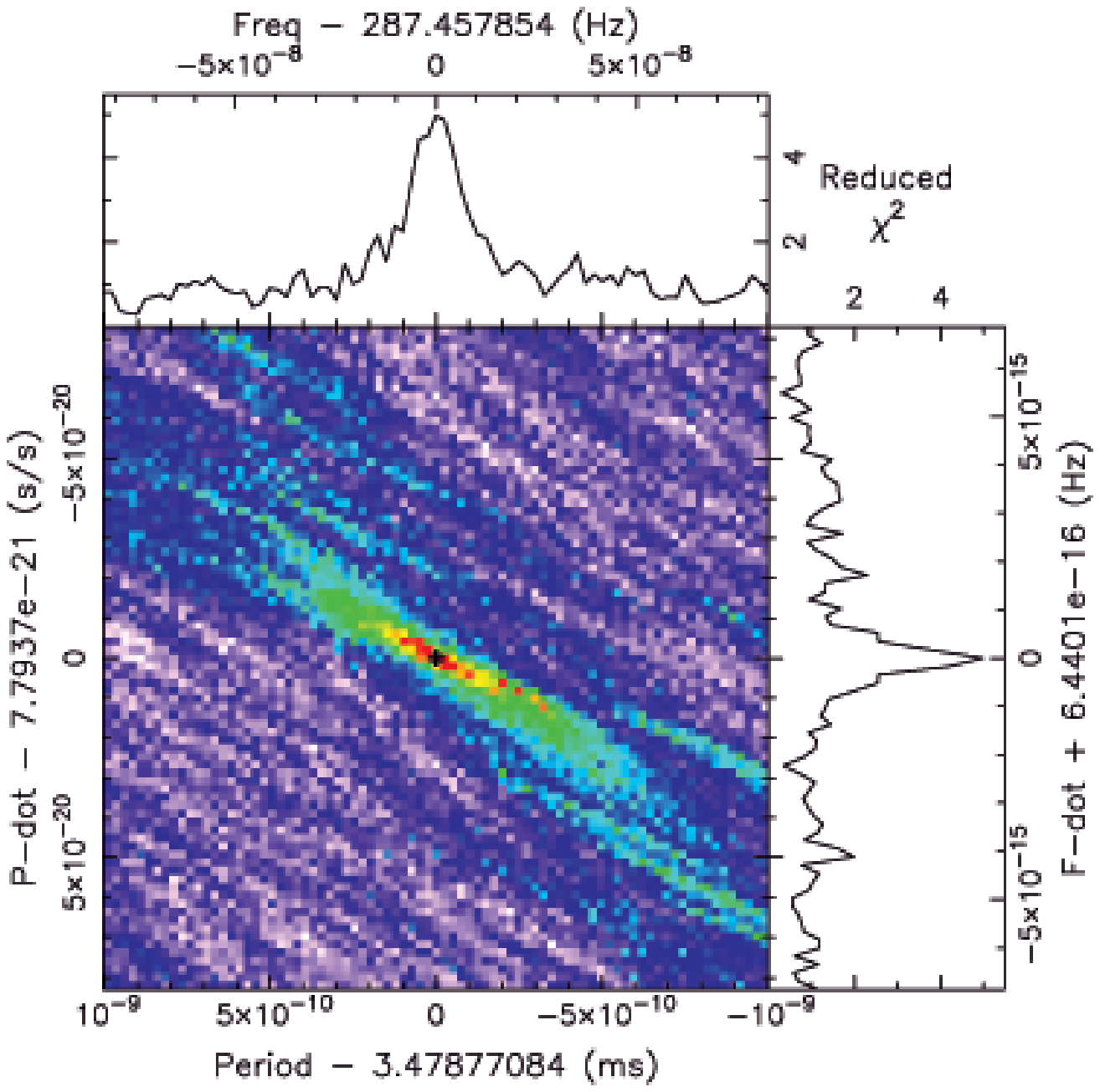,width=8cm,height=6cm}}
\caption{Search for the 3.5\,ms spin period of PSR\,J0751+1807 with {\it PRESTO} in two years of {\em Fermi}-LAT data. {\em Top panel}: here we did not correct the ToAs  for the orbital period, and the pulsar spin period cannot be detected. {\em Bottom panel}: here we correct the ToAs  for the orbital period, detecting a clear signal at the pulsar rotational period.}
\label{msbinary}
\end{figure}


Deep searches for pulsations have been performed in the radio band at several frequencies, with the hope of detecting a fast spinning radio pulsar as in the case of the other TeV binary, PSR\,B1259--63 (Johnston et al. 1999, 2005). However, no radio pulsation has been detected so far from any of these two sources. In that respect it is crucial to note that at periastron PSR\,B1259--63 does not show radio pulsations. Its periastron (given the large orbit, 3.4 year period) has about the same dimension of the major axis of \srca's orbit (26 days period) and it is way larger than \srcb's obit (4 days period), hence it is reasonable to expect that no radio pulsed emission is observed from these much compact binaries.

Before the \CXO\, observations we report here (and in Paper I for \srca), archival observations which could give reliable upper limits on pulsations for a possible fast spinning pulsars ($P \leq 100$\,ms) hosted in \srca\, and \srcb\ were not very constraining, coming mainly from {\it R}XTE\, and {\it XMM--Newton} observations. In particular the high background of these instruments (especially {\it R}XTE) limited the pulsed fraction sensitivity of these observations. For \srca\,  the deepest pulsed fraction limit was derived from a 41\,ks {\it XMM--Newton} observation (Sidoli et al. 2006), which was $<28$\%, in the 12--200\,ms period range. Note that the {\it R}XTE\, monitoring observations performed in 1996 gave an upper limit of only 32\% in the 1--200\,ms range, in fact, even though {\it R}XTE's timing resolution and collecting area was much larger than {\it XMM--Newton}, the much higher background contamination is limiting the sensitivity to weak signals, resulting in a larger pulsed fraction limit\footnote{For that observations, Harrison et al. (2000) claim a limiting pulsed fraction of $\sim$6\%. However they considered the total count rate without correcting for the cosmic and instrumental background, which if corrected increases substantially the upper limit on the detectable pulsed fraction.}. Similarly, for \srcb\, the deepest limits for the presence of a fast pulsar were derived from a 50\,ks {\it R}XTE\, observation performed in 2003, giving an upper limit of the pulsed fraction of $<30$\%. 

We present here the Paper II of our series of \CXO\, observations aimed at searching for pulsed X-ray emission from TeV binaries. In Paper I we presented such analysis for a $\sim$95\,ks \CXO\, Continuous Clocking mode observation of \srca, while in this paper we present a similar analysis on $\sim$70\,ks of \CXO\, data on  \srcb.  In all observations we used the same instrumental set-up, to favor the search for fast spinning pulsars. In \S2 we report on the observation and analysis, we present our results in \S3, followed by a discussion (\S4).


\begin{figure*}
\hbox{
\vbox{
\psfig{figure=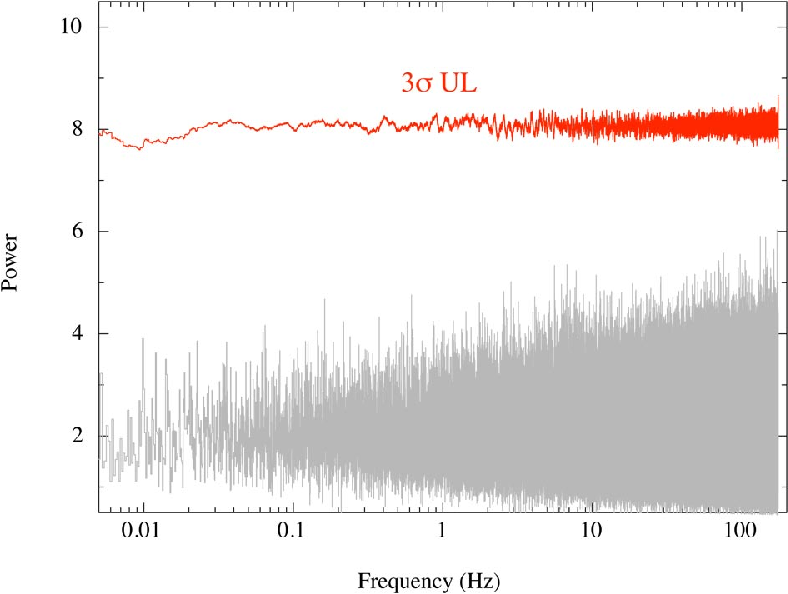,width=8.5cm,height=7cm,angle=270}
\vspace{0.5cm}
\psfig{figure=pf_000285_last.ps,width=8.7cm,height=3.5cm,angle=270}}
\vbox{
\psfig{figure=pf_000285_03_2kev_last.ps,width=8cm,height=3.5cm,angle=270}
\vspace{0.2cm}
\psfig{figure=pf_000285_2_4kev_last.ps,width=8cm,height=3.5cm,angle=270}
\vspace{0.2cm}
\psfig{figure=pf_000285_4_10kev_last.ps,width=8cm,height=3.5cm,angle=270}}}
\caption{{\em Left panels:} Top row: power spectra of \srcb\, in the
  0.3--10\,keV energy band binned at 0.00285\,s with superimposed the
  3$\sigma$ upper limits on the detection of a periodic signal on top
  of the Poisson noise (we used here logarithmic scales to better
  show the low frequency range). Bottom row: pulsed fraction 3$\sigma$
  upper limits on the semi-amplitude of a sinusoidal signal. {\em
    Right panels:} pulsed fraction 3$\sigma$ upper limits versus
  frequency in three different energy ranges (from top to bottom:
  0.3--2\,keV, 2--4\,keV, 4--10\,keV), calculated on the power
  spectrum binned at 0 .00285\,s (left panel) in the 0.005--175\, Hz
  frequency range (see \S\ref{pulsation} for further details).}
\label{dps}
\end{figure*}


\section{Observation and Data analysis}
\label{data}

The Advanced CCD Imaging Spectrometer (ACIS) camera on board of \CXO\,
observed \srcb\, twice, on 2009 July 31st (start time 05:41:49 UT; Obs-ID 10053) and on 2009 August 6th (start time 06:23:48; Obs-ID 10932) for an exposure time of 45.138 and 25.079\,ks, respectively. The observations were performed in Continuos Clocking (CC) mode (FAINT), providing a time resolution of 2.85\,ms and imaging along a single direction. In both pointings the source was positioned in
the back-illuminated ACIS-S3 CCD at the nominal target
position. Standard processing of the data was performed by the {\em
  Chandra X-ray Center} to level 1 and level 2 (processing software DS
ver. 8.0). The data were reprocessed using the CIAO software
(ver. 4.2) and the \CXO\, calibration files (CALDB
ver. 4.2). \\ \indent We corrected the times for the
variable delay due to the spacecraft dithering and telescope flexure,
starting from level 1 data and assuming that all photons were
originally detected at the target position. Furthermore, data were
filtered to exclude hot pixels, bad columns, and possible afterglow
events (residual charge from the interaction of a cosmic ray in the
CCD). Photon arrival times are in TDB and were referred to the
barycenter of the Solar System using the JPL-DE405
ephemeris. 

In order to carry out a timing analysis, we
extracted the events in the 0.3--10\,keV energy range from a region of
5$\times$5 arcseconds around the source position (RA 18:26:15.00, Dec
$-$14:50:53.6). The
source spectrum was extracted from a rectangular region of 5$\times$25
arcseconds around the source position, and the background was taken
independently from a source-free region in the same chip. We extracted
the appropriate response matrix files (RMFs) and ancillary response files (ARFs) for each observation. The source ACIS-S average count rate in the 0.3--10\,keV
energy band was $0.509\pm0.003$\,counts\,s$^{-1}$  and $0.430\pm0.004$\,counts\,s$^{-1}$  for the first and second observation, respectively (see Fig.\,\ref{lcurve}).

\section{Results}
\label{results}

Using the orbital ephemeris of  Aragona et al. (2009), namely an
orbital period of 3.90608$\pm$0.00010\,days and the zero-phase assumed
at T$_0$=52825.485$\pm$0.053\,MJD, we calculated that the first \CXO\,
observation was performed while the compact object was passing from
phase 0.3 to 0.4, while the second observation was spanning phases 0.75--0.9 (see Fig.\,\ref{orbit}).


\begin{figure*}
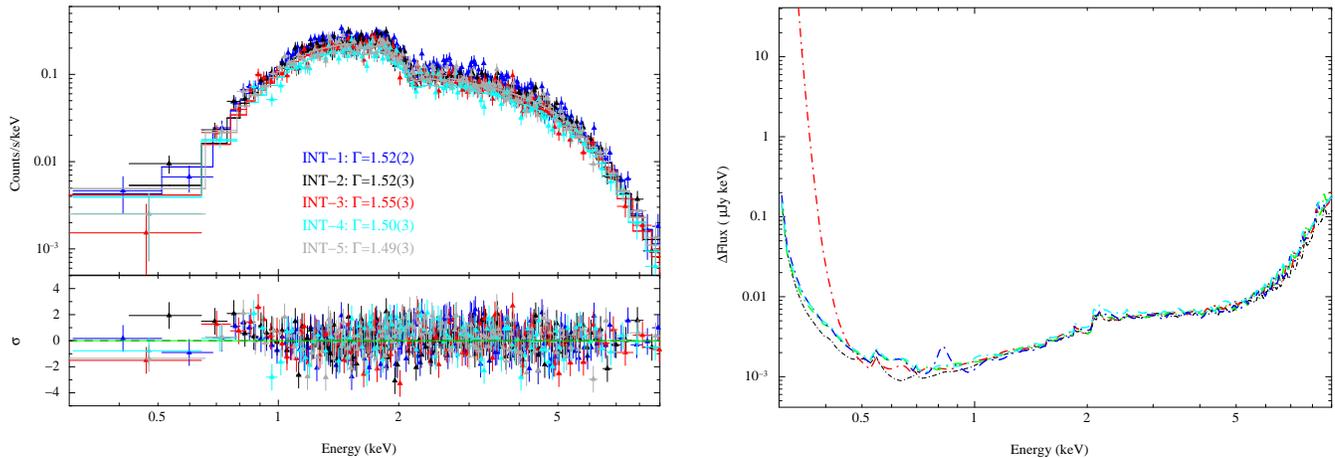

\hbox{
\psfig{figure=spectral_evolution.ps,width=8.7cm,height=6cm,angle=270}
\hspace{0.5cm}
\psfig{figure=insensitivity_plot.ps,width=8.2cm,height=6cm,angle=270}}
\caption{{\em Left panel}:  spectra of the 5 time intervals fitted with an absorbed (N$_{\rm H}=(7.1\pm0.4)\times10^{21}$\cm2) power-law function (see Fig.\,\ref{lcurve} and text for details). {\em Right panel}: 1$\sigma$ sensitivity for the presence of spectral lines (this corresponds to the limit on the flux of detectable spectral lines as broad as the energy-dependent instrument resolution on top of the continuum flux as a function of energy) for each for the five spectra.}
\label{spectra}
\end{figure*}


\subsection{Orbital demodulation}
\label{demodulation}

Given the relatively long \CXO\, exposure time with respect to the $\sim4$\,days orbital period of our target, we had to correct the photon arrival times for the orbital motion of the system, before searching for any periodic signal (which might otherwise be washed out; see e.g. Vaughan et al. 1994 and reference therein).  To correct the arrival times of each X-ray photon for the orbital motion of \srcb\, we  developed an algorithm in the framework of the
program {\tt TEMPO2} (Hobbs et al. 2006). Unfortunately, a dedicated plug-in for the orbit demodulation is not present in the official release, hence we built our own program for this purpose. In particular, we used the {\tt TEMPO2} output format
\texttt{general2}, that calculates the pulsar pulse number and
phase corresponding to each arrival time, giving as input the
orbital ephemeris of a binary system. The algorithm we developed is based
on the creation of a fake {\tt TEMPO2} parameter file that includes the orbital
parameters of  \srcb\, (Aragona et al. 2009), and a pulsar with a constant spin frequency, which we assumed being 1\,Hz.   In this way the 1\,Hz pulsar works like a clock on the reference frame comoving with the true compact object, and the values returned by the output format \texttt{general2} correspond to the photon arrival times in the pulsar reference frame, hence demodulated by the orbital period.

We have tested this algorithm both on simulated and real data. We have first simulated with the {\tt TEMPO2} plug-in \texttt{fake} a 330\,ms pulsar in a binary system with 4 days orbital period. In particular, we simulated the Time Of Arrivals (TOAs) of the pulsar with a Gaussian random noise with a root mean square (r.m.s.) set to 0.1\,ms. We then checked that the residual plot of the simulated TOAs using the full orbital ephemeris we have input was identical to the residual plot of the TOAs  derived demodulating the orbital motion using the above algorithm, when inputting only the pulsar parameters. As a further check, we have applied our demodulation algorithm on two years of {\em Fermi} Large Area $\gamma$-ray Telescope (LAT) data of the binary millisecond pulsar (PSR\,J0751+1807; Nice et al. 2005; Abdo et al. 2009b), which is similarly composed by a 3.5\,ms pulsar in a 0.26\,days orbital period with an helium-rich white dwarf. In Fig.\,\ref{msbinary} we show the {\tt PRESTO} {\tt prepfold} outcome of the {\em Fermi}-LAT data before (top) and after (bottom) applying the demodulation algorithm, showing that after correcting for the orbital parameters we can detected the spin period of the pulsar.

For the orbital parameters we use:  an orbital period of 3.90608$\pm$0.00010\,days, a projected semi-major axis of 3.329$\pm$0.153\,s, a periastron longitude of 236.0$\pm$5.8\,degrees, and an eccentricity of 0.337$\pm$0.036, referred to T$_0$=52825.485$\pm$0.053\,MJD \footnote{For technical reasons, these parameters in Aragona et al. (2009) are referred to an epoch expressed in Heliocentric Julian Day, that differs from the most commonly used Julian Days by a variable amount of time with a maximum difference of 500\,s. Given $\sim$5000\,s error in the epoch determination we neglected the  difference between the Heliocentric JD and the JD we assume in this work.}.

\subsection{Pulsation search and light-curve analysis}
\label{pulsation}

Using the event files with the photon arrival times referred to the Solar System barycenter and corrected for the orbital motion (see above), we searched for periodic and quasi periodic signals with unprecedented
sensitivity, thanks to the extremely low ACIS-S background. Given the length of our observations
(70\,ks in total), the timing resolution of the CC-mode (0.00285\,s), and the number of counts of our observation, we
could search for periodic signals in the $\sim0.005 -175$\,Hz
frequency range. As we explained in detail in Paper I, we
studied the source power spectra performing Fast Fourier Transforms
(FFTs), considering the total exposure time (summing both observations), and on each of the five time intervals shown in Fig.\,\ref{lcurve}. Furthermore, we performed the same search dividing the entire data set in three
energy bands (0.3--2, 2--4, and 4--10\,keV). 

For computing reasons, we performed an average over 13 FFTs with a bin time of 0.00285\,s (Fig.\,\ref{dps} top-left panel), resulting in 2,097,152 bins ($\sim6$\,ks) for each of the 13 averaged power spectra (the averaged power spectrum had a $\chi_{\nu}^{2}$ distribution with 26 degree of freedom (d.o.f.)). For the power spectra produced for the three different energy bands, given the lower number of counts, we could use a single FFT (their power spectrum had then a $\chi_{\nu}^{2}$ distribution with 2 d.o.f.) . We took into account the number of bins searched, and the different d.o.f. of the noise power distribution (Vaughan et al. 1994; Israel \& Stella 1996) in calculating the 3$\sigma$ detection upper limits  reported in Fig.\,\ref{dps} . We accounted for the red-noise (despite negligible in this observation) in the calculation of the detection significance by using a smoothing window technique (see Israel \& Stella 1996 for further details).  We did not find any periodic or quasi-periodic signal.\\
\indent
As reported in Paper I, we computed the 3$\sigma$ upper limits on the sinusoid amplitude pulsed fraction ($PF$), computed according to Vaughan et al. (1994) and Israel \& Stella (1996), which ranges in the 0.3--10\,keV energy band between $PF<$13--21\% (see Fig.\,\ref{dps}  bottom-left panel).  In  Fig.\,\ref{dps} (right column) we calculate the same limits as a function of the energy band, which given the lower number of counts  causes the energy-dependent $PF$ limits to be slightly larger than those derived using the whole datasets.   

In addition, we looked for the presence of SGR--like bursts by binning the counts in intervals of 0.1\,s and searching for excesses above a count threshold corresponding to a chance occurrence of 0.1\% (taking into account the total number of bins), but we did not find any significant short burst. 

In Fig.\,\ref{lcurve}  and \ref{orbit} we report on the light-curve of both observations, which show a variability correlated with the orbital phase of the system, as previously observed in a long {\em Suzaku} observation (Kishishita et al. 2009; Takahashi et al. 2009).


\begin{figure*}
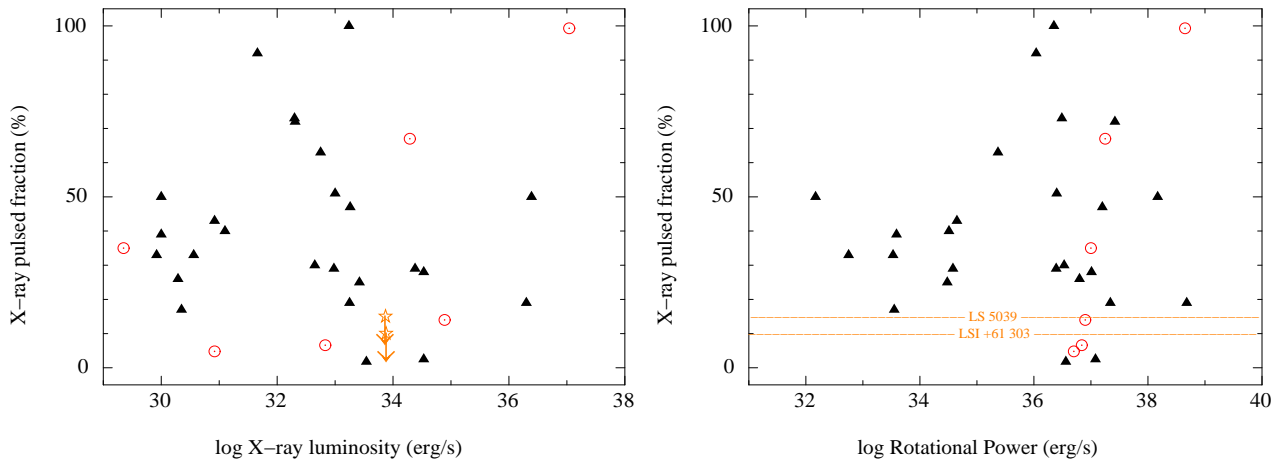

\hbox{
\psfig{figure=pf_lx_psr_final_v2.ps,width=8.5cm,height=6cm,angle=270}
\psfig{figure=pf_edot_psr_final_v2.ps,width=10cm,height=6cm,angle=270}}
\caption{{\em Left panel}:  Pulsed fractions versus total X-ray luminosity. {\em Right panel}: Pulsed fractions versus rotational power. See also Tab.\,1. Circles stand for sources with associated TeV conterparts. Down arrows and horizontal lines refer to the upper limits for \srca\, and \srcb. See \S\,4.2 for details.}
\label{pffigure}
\end{figure*}


\subsection{Spectral analysis: continuum modeling and search for spectral lines}
\label{spectral}

All spectra reported here were binned such to have at least 25 counts per bin and not to over-sample the instrument spectral resolution by more than a factor of 3. We fit the spectra relative to the first and to the second observation with an absorbed power-law, which give a $\chi_{\nu}^2=1.18$ and $\chi_{\nu}^2=1.32$ for 421 and 261 d.o.f., respectively. Given the known spectral variability of \srcb\, as function of orbital phase,  we then searched for a spectral evolution dividing the observation in 5 time intervals (see Fig.\,\ref{lcurve} and \ref{spectra}). Fitting together the spectra of the 5 time-intervals with an absorbed power-law function we found a $\chi_{\nu}^2=1.15$ for 725 d.o.f, with the absorption value tied to be the same for all 5 spectra and the slope and normalization of the power-laws free to vary. We find an absorption value of N$_{\rm H}=(7.1\pm0.4)\times10^{21}$\cm2 (with abundances from Anders \& Grevesse (1989), and cross-setion for photoelectric absorption from Baluchinska-Curch \& McCammon (1998)), and a power-law index which varies only very slightly, from 1.49 to 1.55 depending on the time-interval (see Fig.\, \ref{spectra}; all errors in the spectral values are reported at 1$\sigma$ confidence level). The source 0.5--10\,keV absorbed flux varies from the first to the fifth time interval (from left to right in Fig.\,\ref{lcurve}): $8.8\pm0.2$, $7.8\pm0.2$, $6.7\pm0.2$, $6.0\pm0.2$ and $7.1\pm0.2 \times10^{-12}$\ergscm2 (corresponding to unabsorbed fluxes of $1.2\pm0.2$, $1.0\pm0.2$, $0.9\pm0.2$, $0.8\pm0.2$ and $1.0\pm0.2 \times10^{-11}$\ergscm2 ). 
Furthermore, we searched for possible evidence for  N$_{\rm H}$ variability among the five spectra by means of contour plots with respect to the photon index ($\Gamma$), but while the spectral indexes where found to show a subtle variability (at 2.5$\sigma$ in the best case, between interval 3 and 5; see also Fig.\,\ref{spectra}), we do not have enough counts to constrain any variability in the absorption value.

Comparing our flux and spectral parameters with those derived by {\em Suzaku} at similar orbital phases (Kishishita et al. 2009; Takahashi et al. 2009), we find a perfect match within a 1$\sigma$ error (see Tab.\,2 in Takahashi et al. 2009).

A deep inspection of all spectra did not show evidence for significant absorption or emission features, for which we report in Fig.\,\ref{spectra} (right panel) the 1$\sigma$ limit on the flux of detectable spectral lines as broad as the energy-dependent instrument resolution on top of the continuum flux, as a function of energy (we refer to the {\tt XSPEC} modeling package for details\footnote{http://heasarc.nasa.gov/xanadu/xspec/}).

\subsection{Spatial analysis: search for extended emission}
\label{spatial}
\label{diffuse}

An $\sim$1\arcmin\, extended X-ray emission around \srcb\, has been recently reported by Durant et al. (2011). We produced a one-dimensional radial profile of \srcb\, for both the \CXO\, observations, using the CC mode one-dimensional imaging capability, to search for any evidence for such diffuse extended X-ray emission. We first generated an image of the one-dimensional strip in the 0.3--10\,keV band, and then subtracted the background count rate to remove instrumental and cosmic X-ray background. We extracted a one-dimensional surface brightness distribution extracting photons from 125 annuli centered on the source position and of 2 pixel wide ($\sim$0.99\arcsec). We then did a similar extraction for the simulated {\em Chart/MARX} Point Spread Function (PSF). We did not find evidence for a significant extended emission or structures within $\sim$1\arcmin\, around our target. 

Assuming an absorbed power law  spectrum with N$_{\rm H}=6.4\times10^{21}$\cm2 and $\Gamma=1.9$ for the extended emission between 20--60\arcsec (Durant et al. 2011), we inferred a 3$\sigma$ limiting unabsorbed flux for the presence of X-ray  diffuse emission of $\sim3\times10^{-13}$  in the 0.5--10\,keV energy range. The limit we find is higher than the flux of the X-ray extended emission reported for \srcb\, ($1.4\times10^{-13}$\ergscm2 ; see Tab\,2 in Durant et al. 2011), hence it is compatible with the presence of such emission, and we cannot constrain any further this emission with the observations reported here.

\section{Discussion}
\label{discussion}

We reported here on a $\sim$70\,ks  \CXO\, look at the TeV binary \srcb\, aimed at searching for a dim periodicity in its X-ray emission due to the presence of a pulsar in the system. As in the case of \srca\, (Paper I), we did not find any periodic or non-periodic signal in these observations, which span a significant fraction of the orbit (namely phase 0.3--0.4 and 0.75--0.9). We confirm the orbit-by-orbit stability of the light-curve shape and the modest spectral variations previously observed in this system (Bosch-Ramon et al. 2005; Kishishita et al. 2009; Takahashi et al. 2009).We derived upper limits on the X-ray pulsed fraction of \srcb\, ($PF \lsim15$\%; see \S\ref{pulsation}, and Fig.\,\ref{dps}), as a function of frequency and energy band.

\subsection{On the impact of different ephemeris for \srcb}

A recent analysis of the mass and orbital parameters of \srcb\, was presented by Sarty et al. (2010),  using observations obtained with the MOST satellite. These authors report two main findings: 1) the orbit of \srcb\, has a significantly lower eccentricity than reported before, and 2) the total mass of the binary system is toward the higher end of previous estimates.  
A direct comparison of their results with respect to those of Aragona et al. (2009) and Casares et al. (2005) shows that 
despite the lower mean value of the eccentricity found by Sarty et al. (2010), all of them are consistent within 1$\sigma$ error (see Table 1 in Sarty et al. 2010).  
All of the other orbital parameters are in agreement with those of Aragona et al. (2009) within errors, except for the system mass function, the value of which is instead compatible with that obtained by Casares et al. (2005)\footnote{$f(m)$ (M$_\odot$) =0.0053 $\pm$ 0.0009, 
0.0026 $\pm$ 0.0004, 
0.0049 $\pm$ 0.0006 for Casares et al. (2005), Aragona et al. (2009), and Sarty et al. (2010), respectively.}. For completeness we performed our timing analysis (see \S\,\ref{demodulation} and \S\ref{pulsation}) using the three ephemeris, finding similar results.

\begin{table*}
\begin{minipage}{170mm}
\begin{tabular}{lcccccc}
\hline
Name & log $\dot{E}_{rot}$ & log L$_x$$^{tot}$ & B-field & Period  & PF & Reference  \tabularnewline
 & erg/s & erg/s & G & ms &  & \tabularnewline
 \hline \hline 
B0531+21 (Crab) & 38.65  & 37.04 & 3.78e+12 & 33.40 & 99.3\% & \tiny{Becker et al. 2002; Tennant et al. 2001; Li et al. 2008}   \tabularnewline
B0833-45 (Vela) & 36.84 & 32.83 & 3.38e+12 & 89.29 & 6.6\% & \tiny{Becker et al. 2002; Halpern et al. 2001; Li et al. 2008}   \tabularnewline
B0633+17 (Geminga) & 34.51 & 31.10 & 1.63e+12 & 237.09 & 40\% &   \tiny{Becker et al. 2002; Caraveo et al. 2004}\tabularnewline
B1509-58 (MSH 15-52) & 37.25 & 34.29 & 1.54e+13 & 150.23 &  67\% &  \tiny{Becker et al. 2002; Kuster et al. 2002; Li et al. 2008}\tabularnewline
B1929+10 & 33.59 & 30.00 & 5.18e+11 & 226.51  & 39.0\% &  \tiny{Becker et al. 2002;  S{\l}owikowska et al. 2005}\tabularnewline
PSR B0950+08 & 32.75 &  29.92 & 2.44e+11 & 253.1 &  33.0\%  & \tiny{Zavlin et al. 2004} \tabularnewline
B1821-24 & 36.35 & 33.24 & 2.25e+09 & 3.05 & 100\% &  \tiny{Becker et al. 2002; Rutledge et al. 2004 } \tabularnewline
B0656+14 & 34.58 & 32.98 & 4.66e+12 & 384.87 &  29\% & \tiny{Becker et al. 1997; Chang et al. 2001  }   \tabularnewline
B0540-69 & 38.17 & 36.39 & 4.98e+12 & 50.37  & 50\% & \tiny{Becker et al. 1997; Kaaret et al. 2001}  \tabularnewline
J2124-33 & 33.55 & 30.35 & 3.22e+08 & 4.93 &  17\% &  \tiny{Becker et al. 1997;  Bogdanov et al. 2008  }   \tabularnewline
B1055-52 & 34.48 & 33.42 & 1.09e+12 & 197.10 &  25\% &  \tiny{Becker et al. 2002; De Luca et al. 2005 }  \tabularnewline
J0218+4232 & 35.37 &  32.75 & 4.29e+08 & 2.32 &  63\% &  \tiny{Becker et al. 2002; Kuiper et al. 2002; Mineo et al. 2000 }\tabularnewline
PSR J1617-5055 & 37.20 & 33.26 & 3.10e+12 & 69.33 & 47\%  & \tiny{Becker et al. 2002; Kargaltsev et al. 2009 }  \tabularnewline
PSR J0030+0451 & 33.53 & 30.56 & 2.25e+08 & 4.86 & 33\% &\tiny{Becker et al. 2002 } \tabularnewline
B1937+21 & 36.04 & 31.66 & 4.09e+08 & 1.55 &  92\% &  \tiny{Becker et al. 2002; Nicastro et al. 2002 }  \tabularnewline
J0205+6449 & 37.42 & 32.31 & 3.61e+12 & 65.67 & 72\% & \tiny{Becker et al. 2002; Murray et al. 2002  }\tabularnewline
J2229+6114 & 37.34 & 33.25 & 2.03e+12 & 51.62 &  19\% & \tiny{Becker et al. 2002; Halpern et al. 2001  }  \tabularnewline
J1930+1852 & 37.08 & 34.53 & 1.03e+13 & 136.85 & 2.5\% &  \tiny{Becker et al. 2002; Leahy et al. 2008a; Lu et al. 2007 }  \tabularnewline
J1811-1926 & 36.80 &  30.29 & 1.71e+12 & 64.67 & 26\% &  \tiny{Becker et al. 2002; Torii et al. 1999 }  \tabularnewline
J0537-6909 & 38.68 & 36.30 & 9.25e+11 & 16.11 & 48\% & \tiny{Becker et al. 2002; Marshall et al. 1998, Wang et al. 2001}  \tabularnewline
J1420-6048 & 37.00 & 29.35 & 2.41e+12 & 68.18 & 35\% & \tiny{Becker et al. 2002; Roberts et al. 2001}  \tabularnewline
PSR J1838-0655 & 36.70 & 30.92& 1.89e+12 & 70.5 & 4.8\% &  \tiny{Gotthelf et al. 2008; Lin et al. 2009 }  \tabularnewline
PSR J1846-0258\footnote{Note that PSR J1846-0258 has shown magnetar--like activity (Gavriil et al. 2008; Kumar \& Safi-Harb 2008), hence it should also have magnetic power contributing to its X-ray emission, rather than being "only" rotational powered.} & 36.9 &  34.89 & 4.86e+13 & 325.7 &  14\% & \tiny{Kuiper et al. 2009; Leahy et al. 2008b}  \tabularnewline
PSR J1119-6127 & 36.40 & 33.00 & 4.10e+13 & 407.7 &  51\% &  \tiny{Gonzalez et al. 2005 }\tabularnewline
B0355+54 & 34.65 & 30.92 & 8.39e+11 & 156.38 &   43\% &\tiny{McGowan et al. 2007}  \tabularnewline
B0628-28 & 32.17 & 30.00 & 3.01e+12 & 1244.42 &  50\% & \tiny{Tepedelenl{\i}o\u{g}lu et al. 2005; Li et al. 2008} \tabularnewline
J1124-5916 & 37.07 & 34.53 & 1.02e+13 & 135.31 &   28\% &   \tiny{Hughes et al. 2004; Li et al. 2008}\tabularnewline
B1706-44 & 36.53 & 32.65  & 3.12e+12 & 102.46 & 30\% & \tiny{McGowan et al. 2004; Li et al. 2008} \tabularnewline
J1747-2958 & 36.39  & 34.38 & 2.49e+12 & 98.81 & 29\% & \tiny{Hickox et al. 2009; Li et al. 2008}\tabularnewline
B1951+32 & 36.56 & 33.54 & 4.86e+11 & 39.53 & 1.8\% & \tiny{Chang et al. 1997; Li et al. 2008}   \tabularnewline
J1357-6429  & 36.49 &  32.30 & 7.83e+12 & 166.1 &  73\% & \tiny{Zavlin et al. 2007} \tabularnewline
\hline
\hline
\end{tabular}
\caption{Parameters for all isolated rotational-powered pulsars with detected X-ray pulsations. 
The  $\dot{E}_{\rm rot}$ is the rotational power as derived from the rotational period and its derivative; L$_x$$^{tot}$ is the total X-ray luminosity, and the PF is defined here as $\frac{N_{max}-N_{min}}{N_{max}+N_{min}}$, where N$_{max}$ and N$_{min}$ are the maximum and minimum number of counts in the folded light-curve. We refer to the quoted papers for more details on L$_x$$^{tot}$ and PF, which were always in an energy range close to 0.3--10keV.  The values of the the rotational power ($\dot{E}_{rot}$), B and the period were all derived from the ATNF pulsar catalog (Manchester et al. 2005). See also Fig.\,\ref{pffigure}.}
\end{minipage}
\label{pftable}
\end{table*}


\subsection{Comparison of our limits with X-ray pulsed fractions of known pulsars}
\label{pulsars}

We use the deep $PF$ upper limits we derived here for \srcb\, (see \S\ref{results}, and Fig.\,\ref{dps}), and in Paper I for \srca, to draw a comparison between the X-ray emission of the putative pulsar in these TeV binary systems, with that of accreting pulsars in HMXBs, and isolated rotational-powered pulsars. In the HMXB case, after a thorough search in the literature, the large variability of such pulsed fractions (from a few \% to 90\%) as a function of luminosity and energy even in a single binary system, did not allow us to draw any confident conclusion from the comparison with our limits. The large variability of the PF in these systems is probably due to the variable accretion rate, and accretion column geometry. 

On the other hand, in Tab.\,\ref{pftable} and Fig.\,\ref{pffigure} we report on the comparison between the X-ray PF values of all isolated rotational-powered pulsars where this value has been measured, and the limits we have for \srca\, and \srcb. All pulsed fractions here are calculated as:  $$\frac{N_{max}-N_{min}}{N_{max}+N_{min}}$$ where $N_{max}$ and $N_{min}$ are the maximum and minimum number of counts in the X-ray folded light-curve, respectively. We have tried to use the same X-ray band for all sources, although not for all of them the PF value was reported in the 0.3-10\,keV energy range we wanted. In these few cases  where only the 0.3--2\,keV or 2--20\,keV range were available (as e.g. for J1930+1852 which was observed by \RXTE), we have checked that the PF values had only a marginal dependence with energy, and assumed that the same value did hold also for the 0.3--10\.keV energy range. Furthermore, for those sources for which in the original paper the PF was given using a different definition, we have extracted the values from the pulse profile directly, such to compare PF values derived as coherently as possible.  In Fig.\,\ref{pffigure} we plot the values reported in Tab.\,\ref{pftable}, using red circles for pulsars being associated with a TeV source (see also Mattana et al. 2009), and black triangles for pulsars with no detected TeV counterpart. Note that the TeV emission is mainly related to the presence of a pulsar wind nebula sustained by the energetic pulsar. 

The luminosities we report in  Fig.\,\ref{pffigure} for \srca\, and \srcb\, are 7.6 and $7.5\times10^{33}$\ergs\, (they assume a distance of 2.3\,kpc and 2.5\,kpc, respectively), and were derived from the average source flux reported in this work and Rea et al. (2010, Paper I), hence they refer to the orbital phase ranges where such observations were performed.

At variance with what was stated in Paper I, although it is true that isolated rotational powered pulsars have on average PF larger than the limits we found for TeV binaries, the current detailed analysis on their X-ray pulsed fractions using a coherent definition of $\frac{N_{max}-N_{min}}{N_{max}+N_{min}}$ for all pulsars, does not seem to show a trend of higher PF values for sources with TeV counterparts.

\vspace{1cm}
\noindent
This research has made use of data from the Chandra X-ray Observatory and software provided by the Chandra X-ray Center. We thank G.L. Israel for useful suggestions, and for allowing the use of his {\tt DPS} software. We thank S. Zhang, J. Li and Y. Chen for discussion, and the anonymous referee for valuable suggestions. NR is supported by a Ramon y Cajal Research Fellowship to CSIC. This work has been supported by grants AYA2009-07391, SGR2009-811 and TW2010005.

\label{lastpage}

\end{document}